# Tracking Systems as Thinging Machine: A Case Study of a Service Company

Sabah S. Al-Fedaghi, Yousef Atiyah
Computer Engineering Department
Kuwait University, Kuwait

***Abstract*—Object tracking systems play important roles in tracking moving objects and overcoming problems such as safety, security and other location-related applications. Problems arise from the difficulties in creating a well-defined and understandable description of tracking systems. Nowadays, describing such processes results in fragmental representation that most of the time leads to difficulties creating documentation. Additionally, once learned by assigned personnel, repeated tasks result in them continuing on autopilot in a way that often degrades their effectiveness. This paper proposes the modeling of tracking systems in terms of a new diagrammatic methodology to produce engineering-like schemata. The resultant diagrams can be used in documentation, explanation, communication, education and control.**

## I. Introduction

Transportation is a crucial element in modern society. Transportation, here, refers to the physical movement of things. The Global Positioning System (GPS) plays important roles in tracking the movement of things and overcoming problems like safety, security and other location-related applications. From one perspective, it is claimed that GPS is "one of the most important inventions in the last 25 years" [1]. The use of GPS devices in positioning and tracking things has increased immensely in the past decades, and applications include mobility pattern recognition, vehicle navigation, fleet management and route tracking [2].

GPS technology is still relatively new, which raises many issues for potential users. It was originally designed for military application, and it combines satellite navigation systems that broadcast location information (e.g., latitude and longitude, speed, heading and altitude) across the Earth [2]. GPS usually requires at least four satellites to be on the visible horizon.

The focus of this paper is on vehicle-tracking systems. A typical system includes a mechanized device and software at an operational base to locate and monitor the position, timing and mobility of a vehicle. It utilizes a GPS receiver along with a Global System for Mobile Communications (GSM) modem to monitor the vehicle status in terms of time and position. GSM has become an accepted worldwide standard. Transmitting data in a GSM-supported modem is achieved by utilizing radio signals.

The base station is supplemented by products such as Google Maps and Bing Maps that provide maps using satellite imaging that shows objects indicating points of interest or line objects to show tracks. The interaction between a GSM modem and a GPS receiver is facilitated by a microcontroller [3].

The aim of this paper is to utilize a new diagrammatic language in producing a conceptual (non-technical) model of vehicle-tracking systems.

### A. Conceptual Modeling

Thinking diagrammatically as a way of conceptualizing our world has been in existence from the moment the first cave-person picked up a soft "rock" and started making markings on the walls of his/her dwelling. As civilization progressed, humanity moved into recording our activities and learning via tablets, papyri and paper for posterity [4].

Modeling can take the form of abstraction, idealization and representations of what is observable from nature [4]. In software engineering, thing-oriented modeling is used to model a portion of reality through flow (abstract) machines [5]. We will utilize such a methodology in developing the conceptual model of vehicle-tracking systems.

### B. Problem: Unsystemic Process Description

Problems of describing processes such as vehicle tracking arise from reported difficulties in creating a well-defined and understandable model of such processes [6]. Modeling is a way to visualize processes to give maximum efficiency, regardless of complexity [7]. Currently, processes are fragmental, nameless and invisible phenomena that most of the time result in difficulties creating documented descriptions [8]. Additionally, most organizations have repeated tasks that, once learned, continue on autopilot and, as they are passed on from one employee to another, often degrade in their effectiveness. "These businesses seldom diagram their processes, as they often lack the knowledge, understanding and willingness to invest the time and effort, or a combination of these" [9]. However, different types of unsystemic (neither uniformed nor holistic in the sense of applying to parts and the whole of a system) diagrams are utilized in process descriptions.

The following provides sample diagrams used in detailing tracking systems. Bharati and Fernandes [3] described a tracking system where a microcontroller is linked to the GPS. The data are received by the GPS receiver from satellites, and data are processed and sent to the GSM modem. On the user end, a GSM-enabled device acts as the GSM modem connected serially to the microcontroller. Such a process is depicted in Fig. 1. Verma et al. [10] used flowcharts in describing their tracking system, as shown in Fig. 2. Manual pictures (e.g., satellite images, a human figure on a computer screen, a car or a network tower) are also used in explaining the architecture of a GPS vehicle-tracking system (see [11]). ER, UML and DFD are also used in this context (e.g., see Fig. 3 [12]). Benrouyne [13] utilized UML use cases. We will model use cases using our diagrammatic language in section 3.

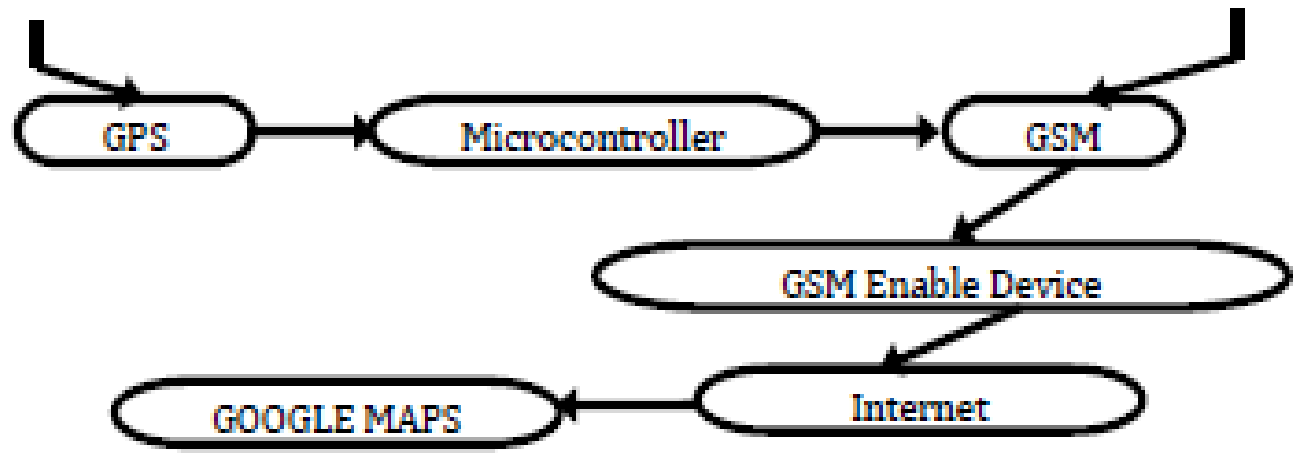

Fig. 1. Block Diagram Illustrating the Concept (Adapted from [3]).

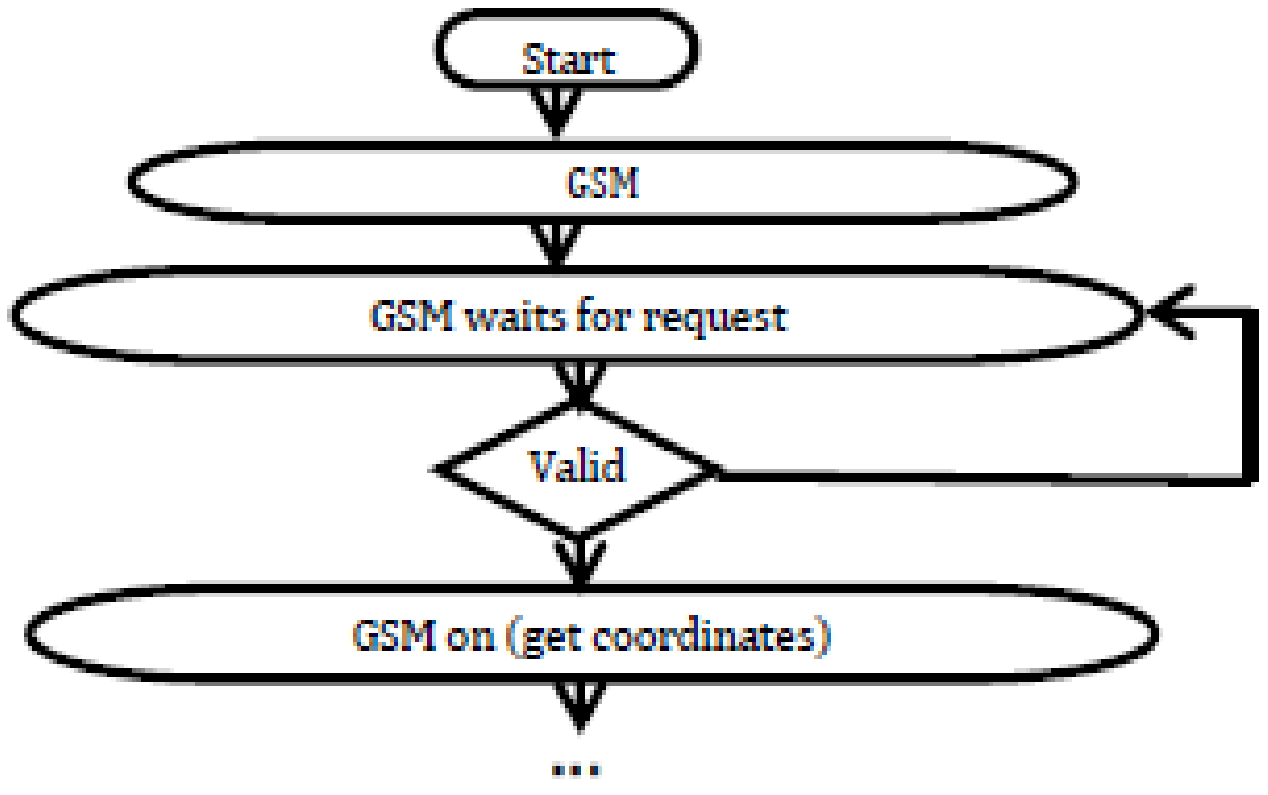

Fig. 2. Flowchart of a Tracking System (Adapted from [10]).

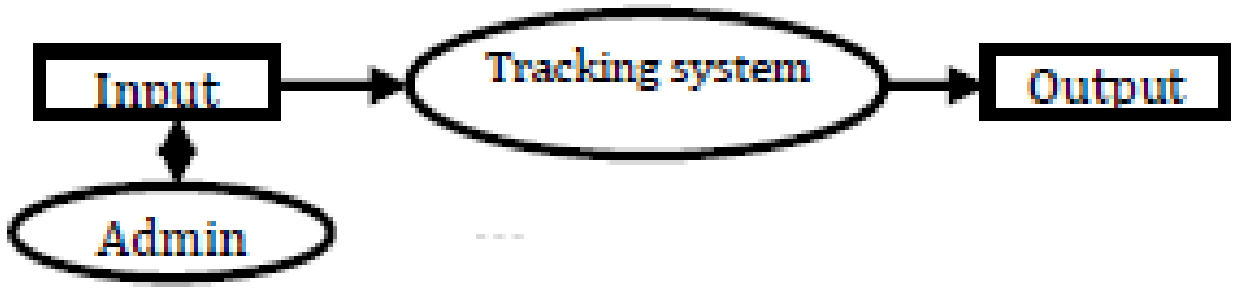

Fig. 3. Level-0 of Describing a Tracking System (Adapted from [12]).

### C. *Proposed Method of Description*

This paper proposes modeling tracking systems in terms of a new methodology that diagrams such a process to produce engineering-like schematization. The resultant illustrations can be used in documentation, explanation, education and control. Specifically, we focus on vehicle-tracking systems that aim at locating and monitoring vehicles, thus enabling an agency to observe activities related to its fleet of vehicles (e.g., sending alerts, providing security and finding paths to a particular destination). We utilize a new diagrammatic language, called the thinging machine (TM)—also called the flowthing machine (FM)—which will be reviewed briefly in the next section [14-20] as a foundation for describing a tracking system. The limited number of notions used in the TM and its all-inclusive description that encompasses the cyber–physical system make TM attractive for modeling tracking systems. Section 3 applied this methodology to an example in the literature that uses UML for this purpose. Section 4 discusses the notion of tracking and monitoring in the context of TM. Section 5 develops a TM model for an operational tracking system in Kuwait.

## II. Thinging Machine (TM)

Drawing on Deleuze and Guattari [21], Bryant [22] declared, “All objects can be understood as machines.” TM modeling utilizes an *abstract thinging machine* (hereafter, *machine*) with five stages of thinging, as shown diagrammatically in Fig. 4.

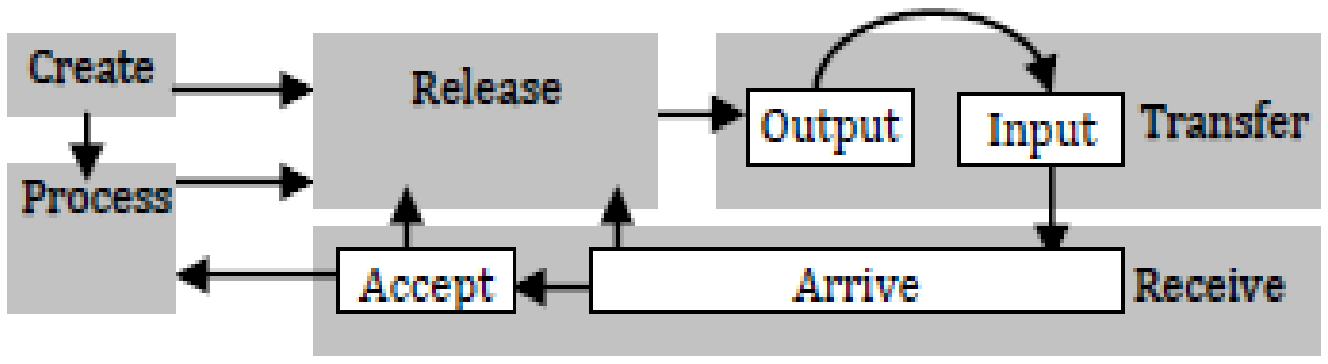

Fig. 4. Thinging Machine.

In philosophy, *thinging* refers to “defining a boundary around some portion of reality, separating it from everything else, and then labeling that portion of reality with a name” [23]. TM modifies the notion of thinging attributed to Heidegger [24] through applying it to the life cycle of a thing and not to its ontological feature. In TM, a thing is created, processed, released, received and transferred as will be described next. Moreover, in TM, a thing may not have a name as mentioned above, but it is distinguishable in some way. This distinguishability installs it as a thing in a system (machine). For example, the thing we now call *oxygen* had been described (created in the human knowledge system) prior to its discovery in 1774 when Priestley called it “dephlogisticated air” and did not recognize it as a chemical element. The name *oxygen* was given in 1777 as a chemical element in the chemical system.

In TM, a system (machine) has (created in it) a thing only if it “knows” (becomes aware or experiences) the thing. This is a phenomenological view projected over systems. In phenomenology, things are given to us through sight, touch, taste or smell, but in TM things are given *to the system* by machines, otherwise the things do not appear at all (in the system). In TM, the system is a grand machine—a machine that is not contained in a machine. *Creation* in a machine is not a thing, but rather a process that provides the machine (and the system) with new things.

According to Heidegger [24], to understand the thingness of things, one needs to reflect on how thinging expresses the way a “thing *things*” (i.e., “gathering,” uniting or tying together its constituents, just as a bridge makes an environment [banks, stream and landscape] into a unified whole). From slightly different perspectives, saying thinging and things *thing* (verb) refers to actualization (manifestation), existence, being known or recognized, possession of being, being present, being there, being an entity (a creature), appearance or the opposite of nothingness. According to [25], Heidegger’s view can however be seen as a tentative way of examining the nature of entities, a way that can make sense. An artefact that is manufactured instrumentally, without social objectives or considering material/spatial agency, may have different qualities than a space or artefact produced under the opposite circumstances.

In TM, a strong association exists between systems and their models. A system is defined through a model. We view a system as an assemblage of things and machines. In simple words, as will be exemplified later, it is a web of (abstract) machines represented as a diagram (the grand machine). A machine can *thing* (i.e., creates, process, receive, transfer and/or release other things). These operations (i.e., thing or create, process, receive, transfer and release) are represented within an abstract Thinging Machine (TM) as shown in Fig. 5.

The starting phase of development in a system's life cycle includes collecting information in order to construct a conceptualization blueprint of it. This conceptualization is intended to give an adequate description of the system (machine) boundaries so that needs, scope and constraints are taken into consideration in building the overall system description that conceives it in its entirety. This paper diagrammatically describes a sample tracking system based on the notion of the verbs *to thing* and *thinging*.

A machine things (verb): (i.e., creates, processes, receives, releases and transfers things). It handles things and is itself a thing that is handled by other machines. The TM model is a grand thing/machine that forms the thinging of a system. Thinging here refers to the creation, processing, receiving, releasing and/or transferring of the system (grand machine) or any of its sub-machines.

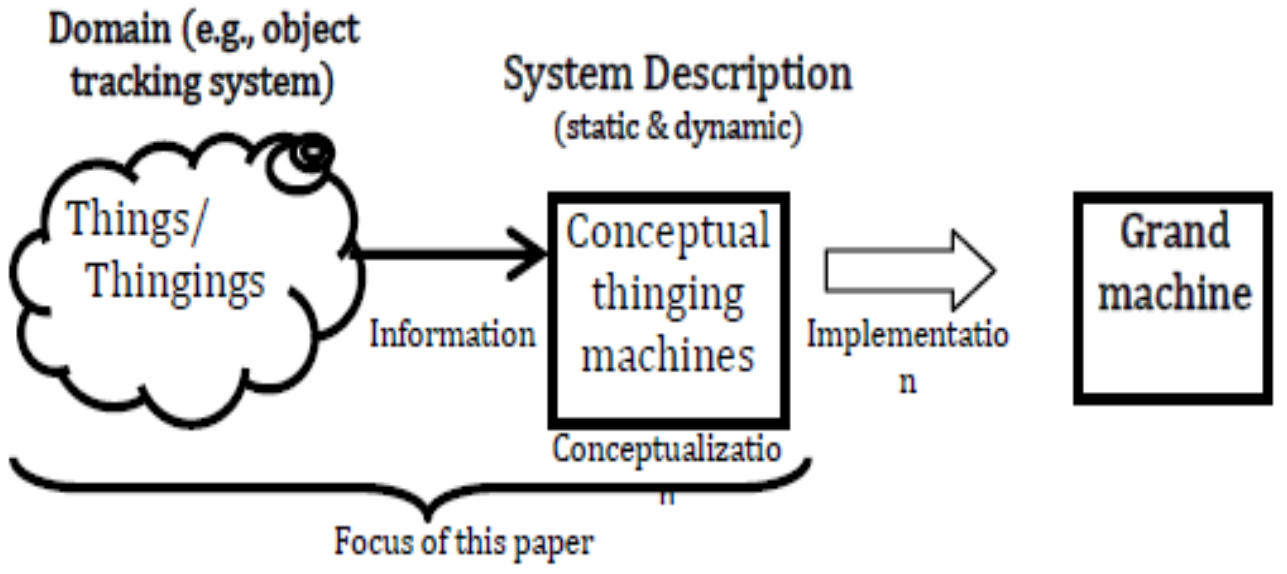

Fig. 5. The Aim is to Capture Things/Thinging in the Form of a Grand Diagram.

The stages in the machine can be briefly described as follows.

**Arrive:** A thing flows to a new machine (e.g., packets arrive at a buffer in a router).

**Accept**: A thing enters a machine; for simplification purposes, we assume that all arriving things are accepted; hence, we can combine Arrive and Accept into a **Receive** stage.

**Release**: A thing is marked as ready to be transferred outside the machine (e.g., in an airport, passengers wait to board after passport clearance).

**Process** (change): A thing changes its form but not its identity (e.g., a number changes from binary to hexadecimal).

**Create**: A new thing is born in a machine (e.g., a logic deduction system deduces a conclusion).

**Transfer**: A thing is inputted or outputted in/out of a machine.

TM includes one additional notation, triggering (denoted by dashed arrow), that initiates one flow from another.

## III. Example

Benrouyne [13] used UML in the analysis phase to identify and organize the requirements of an anti-theft vehicle-tracking system. Fig. 6 represents the use case diagram of the different actors of the system including the end user, the administrator and the GPS tracker. Some of the use cases are given as follows [13]:

*1)* The user or administrator enters the login username and password.

*2)* The user or administrator is logged in the system.

*3)* The user starts this use case by clicking on *create account* on the menu.

*4)* The user provides a valid International Mobile Equipment Identity (IMEI) of a GPS tracker.

*5)* The user fills the required fields of the creation form of the desired account.

*6)* The end user selects a view of the vehicle's current location.

*7)* Live positioning of the vehicle is displayed as the vehicle moves.

*8)* The user enters an alert radius.

*9)* The user selects the "enable proximity alerts" button.

*10)* The proximity alert is set and enabled.

*11)* A confirmation message pops up.

*12)* The user presses "confirm" to enable proximity alters.

*13)* An alert radius is displayed on the map.

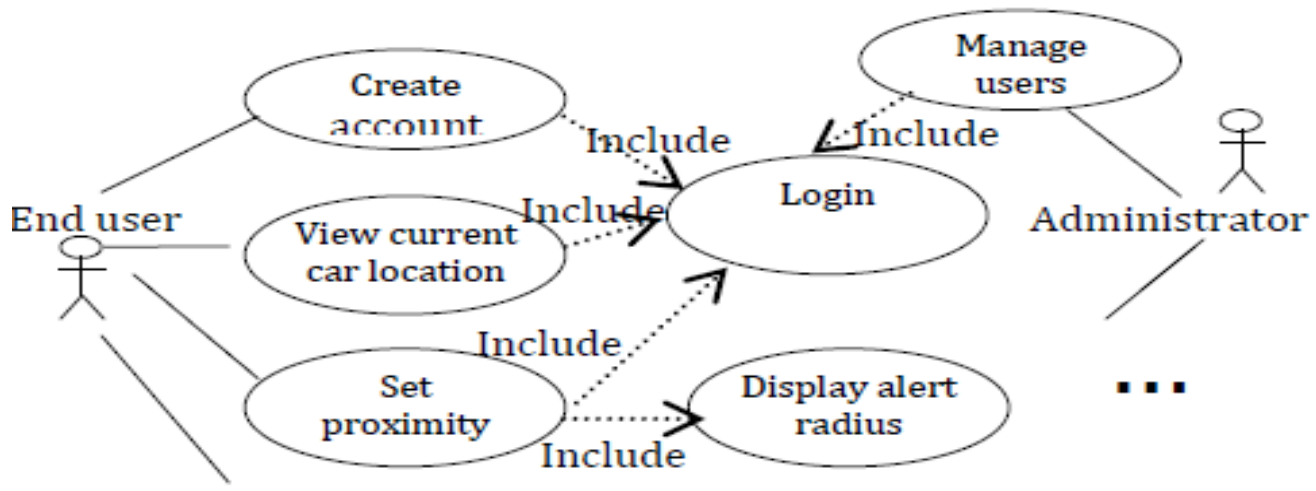

Fig. 6. Use Cases (Adapted from [13]).

### D. Static Representation

Fig. 7 shows the TM static representation of some of these use cases. For simplicity sake, boxes are removed. In Fig. 7, first the user creates a request to login supported by the account name and password (circle 1). The request flows (2) to the system where it is processed. If it is valid (3), then a session is opened (4) and displayed to the user (5). The user requests an account for the tracking system (6). After processing the request (7), the system asks the user to send the IMEI and other requested information (8). The user sends the requested data (9). The system processes the standard vehicle-tracking menu (interface; 10) and downloads it to the user (11).

Now the user is able to use the vehicle-tracking system. He/she selects from a menu (12; e.g., current vehicle location), and such a choice flows to the system where the current vehicle location is processed (13) and triggers the creation of a map (14) that flows to the user (15).

We assume that the user has the tracking system menu available to him/her (16). He/she opts to create an alert radius. This is transferred to the system to be processed (17) and causes the creation of the alert (18) and setting it (20), and then a confirmation message is sent to the user (21). The user signals that the alert is enabled (22), and processing this signal creates an alert map (23).

## E. *Dynamic Representation*

Now, using the static description in Fig. 7, we identify different meaningful events to be used to build a certain sequence of events. An event in TM is a machine that includes at least three submachines: time, region and the event itself. For example, the *log in to the system* event is modeled as shown in Fig. 8. The region is the space where the event occurs (sub-diagram of the static description in Fig. 7). Accordingly, we identify the following events (see Fig. 9) in the static description as follows:

Event 1 ($E_1$): Logging in with username and password.

Event 2 ($E_2$): Opening a session.

Event 3 ($E_3$): Requesting an account in the vehicle-tracking system.

Event 4 ($E_4$): Requesting IMEI and other information.

Event 5 ($E_5$): IMEI and other information are received.

Event 6 ($E_6$): Displaying the tracking system menu.

Event 7 ($E_7$): Selecting current vehicle location.

Event 8 ($E_8$): Finding the latest vehicle coordinates and constructing a map that is displayed.

Event 9 ($E_9$): Selecting alert radius.

Event 10 ($E_{10}$): Creating proximity alert, setting it and sending a confirmation message.

Event 11 ($E_{11}$): The alert is enabled.

Event 12 ($E_{12}$): Alert map is displayed.

We assume that the events are not independent of each other; hence, Fig. 10 shows the chronology of these events.

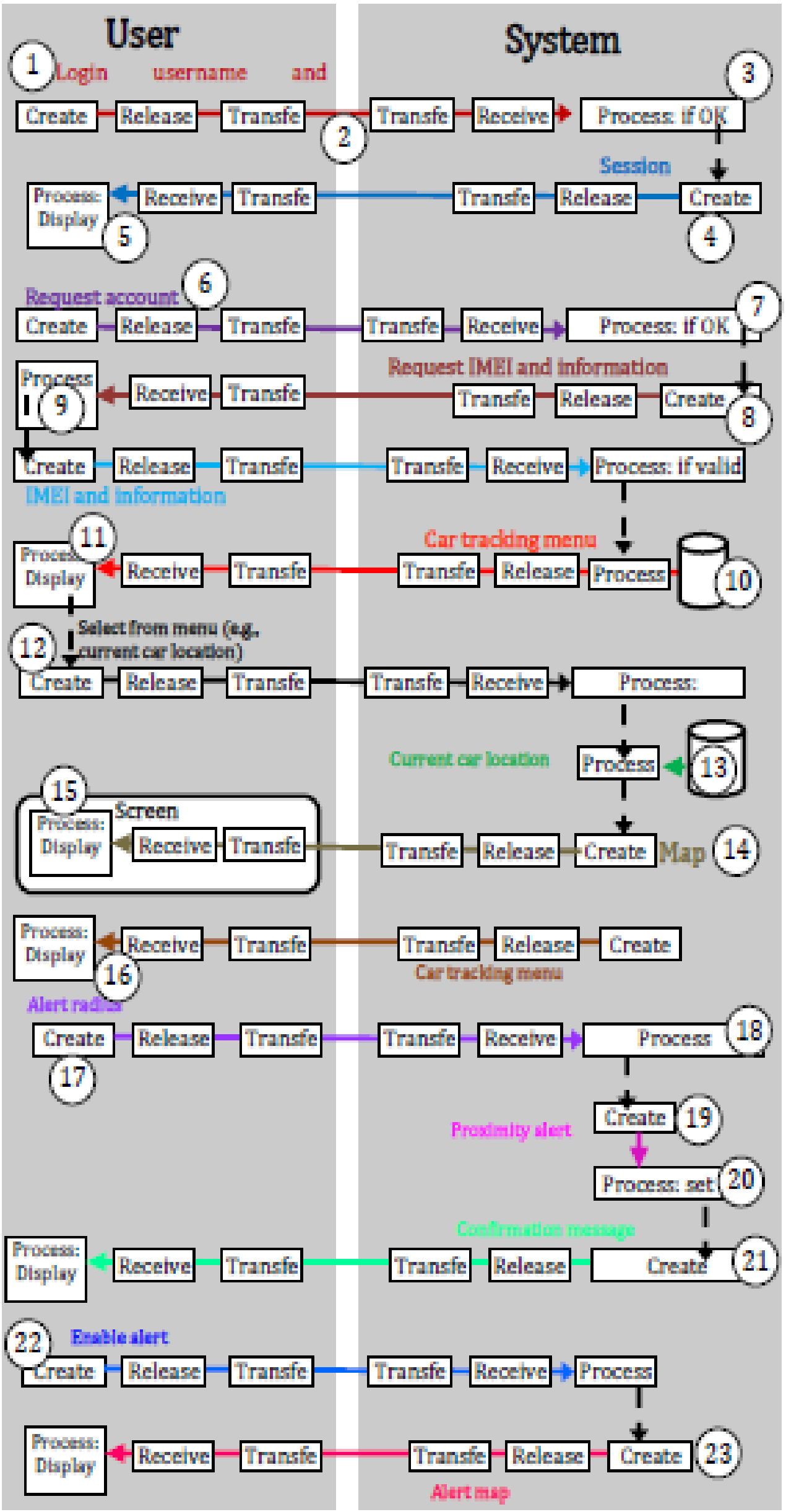

Fig. 7. Partial TM Representation of an Anti-Theft Car-Tracking System.

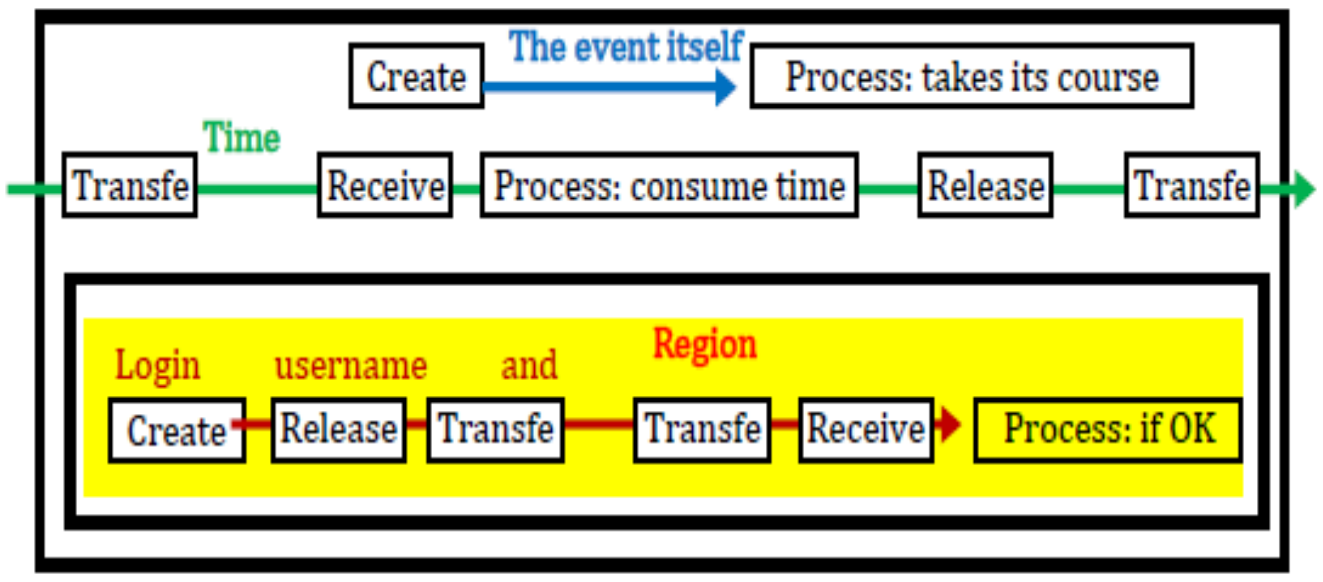

Fig. 8. Description of Event Machine.

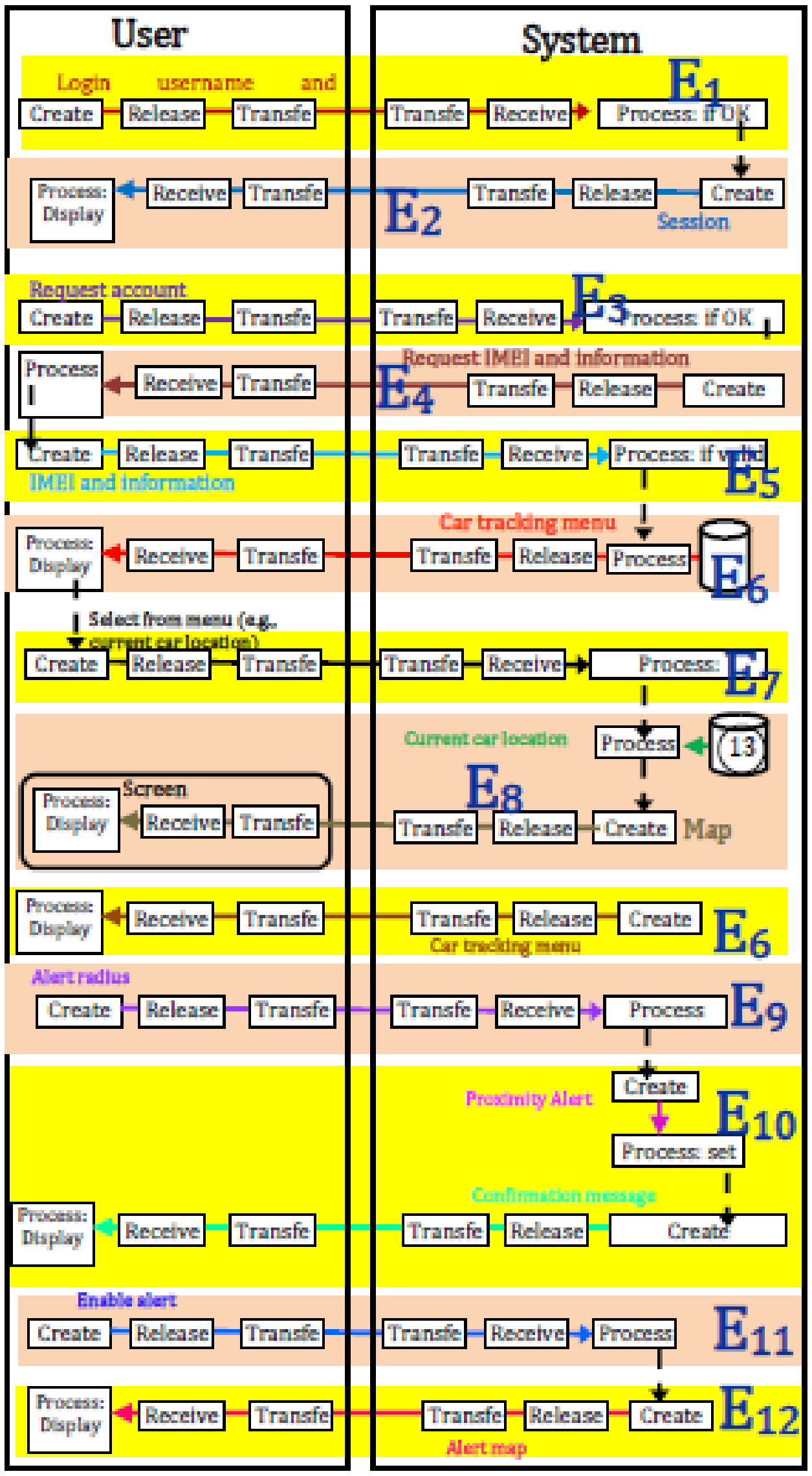

Fig. 9. Events on the static description of the tracking system.

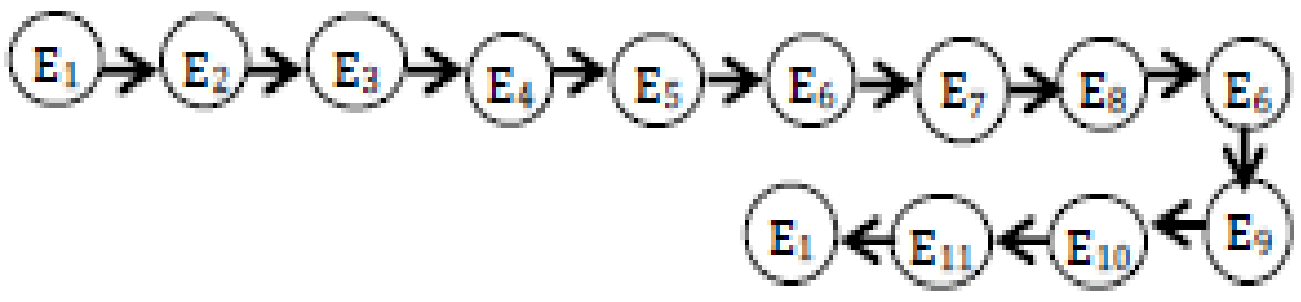

Fig. 10. The chronology of events in Fig. 9.

## IV. Tracking: Definition and Review

The dictionary meaning of *tracking* refers to following the trail or movements of something, typically to find it or note its course. Tracking may also be capable of providing monitoring. Monitoring denotes collecting and processing data about a system to check the progress of something over a period of time. A tracking system involves collecting, processing, aggregating and displaying real-time quantitative data.

Tracking is used in monitoring enterprise elements that are processed and presented in an understandable format. Such a reporting process helps an administrator understand the performance, current status of work and what is normal in the enterprise. With data from reports, an administrator can make informed decisions for capacity planning, maintenance, troubleshooting and security. Although reporting helps one to understand what is normal and the current status of an enterprise, alerts help administrators to identify possible issues related to performance and security and to assess potential problems. All of this is applied to tracking systems.

Fig. 11 shows a general picture of tracking and monitoring from the TM point of view. Feedback can easily be added to such a picture to facilitate control. This is also applied to any tracking system.

A great deal of the research and a body of commercial studies report the classification and limitations of tracking systems (see surveys in [26-28]). These studies include methodological studies that evaluate the reliability of GPS-based traffic data, technical studies that outline the general procedure for processing GPS tracking data and related issues and practical studies that investigate the influence of road network conditions (e.g., density) to find an optimal location. Specifically, since the 1990s, many have studied the routes of vehicles [26] [29-30], the travel pattern and prediction of human mobility [31-32] and environment control [33]. Leduc [34] discussed the potentials related to new GPS technologies. Spek et al. [35] examined the GPS potential applications of spatial-temporal data, thus adding new knowledge to urban studies. Additionally, considerable effort has been involved in processing and analyzing collected data (e.g., achieving precision instantaneously) [2]. In this paper, we are not concerned with the data aspect of a vehicle-tracking system; rather, we focus on the issue of providing visualization of the entire system.

Accordingly, no literature directly relates to our study, but some works have used diagrammatic languages such as UML in describing systems without explicating targeting the modeling topic. To demonstrate the feasibility of applying TM in a tracking system, we next model an actual tracking company in Kuwait.

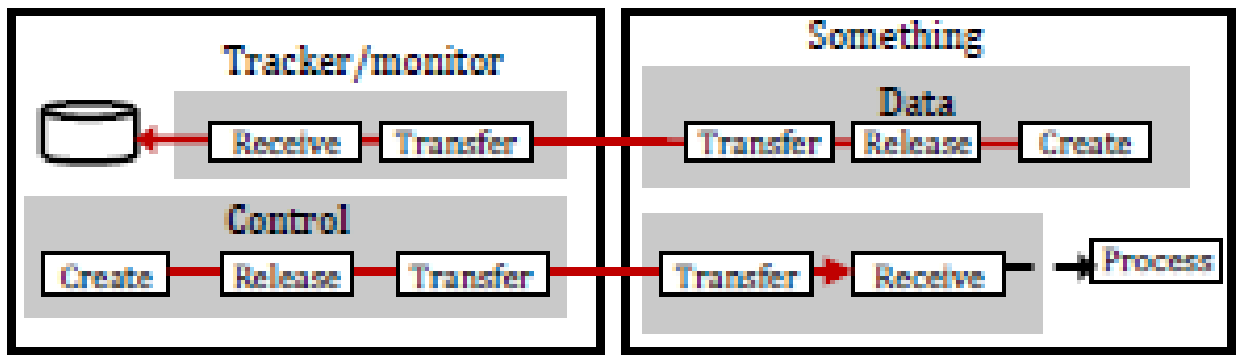

Fig. 11. TM Definition of a Tracking/Monitoring/Control System.

## V. Case Study in Kuwait

Currently, any tracking process in Kuwait is monitored by a local authority known as the Communication and Information Technology Regulatory Authority (CITRA). Kuwait law prohibits private installation of vehicle trackers due to privacy issues; therefore, any installation of GPS trackers should be on a corporate level. Any provider of tracking services must register with CITRA. Every imported device must also be registered in CITRA records.

The current problem is that no clear understanding and standard documentation and regulations exist for the tracking process. This makes it difficult for service providers to present and manage the process for multiple clients with different needs and requirements. It creates an obstacle to fulfilling the utilization and growth of such services.

We take a current working company as a case study from the point of view of how the company describes and documents its tracking system. Of course, the implication is that TM can be used for any tracking system in the world.

*A. Description of Current System*

The current system of tracking vehicles requires installing a tracking device within each vehicle and an underlying infrastructure of servers. The system involves the following:

- A black box tracking device that is installed inside the vehicle to be tracked;
- A tracking server that is responsible for managing and communicating with the device;
- A tracking system that contains an interface to view data and to control the device.

No official documentation of this process exits, but it is understood by the workers in the company. Accordingly, a new employee receives an oral description of the system during his/her training period.

*B. TM Model of the Static Model*

Fig. 12 shows the current system of tracking a fleet of vehicles by installing a tracking device within each vehicle. It is divided into sub-processes as follows:

*1) Gathering data from the vehicle*: In the diagram in Fig. 12, vehicle data flows from the vehicle (circle 1 in the diagram) to the tracking device (2) and includes the following:

Ignition Status: The engine is ON or OFF (3).

Speed: The current speed (4).

Temperature: The current temperature (5).

Acceleration: The current acceleration (6).

All data flow to become part of a message that is sent to the tracking server, as will be described.

*2) Acquiring GPS location*: The tracking device simultaneously receives data from four satellites using an antenna (7) that obtains navigation data from the satellites (8) through a receptor (9). The data are processed (10) to trigger (11) the generation of the coordinates (12) of the location of the vehicle that flow to become part of the message.

Fig. 12 may look complex. However, note that the TM description forms the base for different levels of representation of the system. For example, if we want to discuss different components of the system and the flow among them regardless of the role of a machine (components)—whether it generates, processes or transports things—we can remove the stages of machines, thus cutting the size of the diagram by more than half of its size, as shown in Fig. 13.

Continuing with the description in Fig. 12, the next process is as follows:

*3) Data processing and message creation*: The tracking device receives the readings (i.e., (3), (4), (5) and (6) in Fig. 12) from the vehicle and from the antenna (12), then the device records its clock time (13) and its device ID (14) and all these data are processed (15) to construct a message (16). After saving a copy of the message (17), the message flows to the tracking server (18).

*4) Message entering the tracking server*: The message is transferred via a GPRS connection to the tracking server (18) where a copy is stored (19), and the message is processed in the server (20) to create a formatted message (21) that flows to the tracking system (22). The tracking system is a web application (website) that is accessed by users/operators though a web browser and managed by the tracking server.

*5) Message processing in the tracking system*: In the tracking system, the message is defragmented to its original fields of data: temperature (23), speed (24), acceleration (25), clock time (26), ignition status (27), coordinates (28) and ID (29). Each type of data is handled independently by the tracking system as follows:

*a) Generate alerts: If*

- Temperature is outside two fixed temperature ranges (30),
- Speed exceeds fixed speed limit (31), and/or
- Acceleration exceeds a fixed G-Force value (32).
- Then, corresponding alerts (33), (34) and (35) are created and displayed on the interface (36) and transferred to the driver (37).

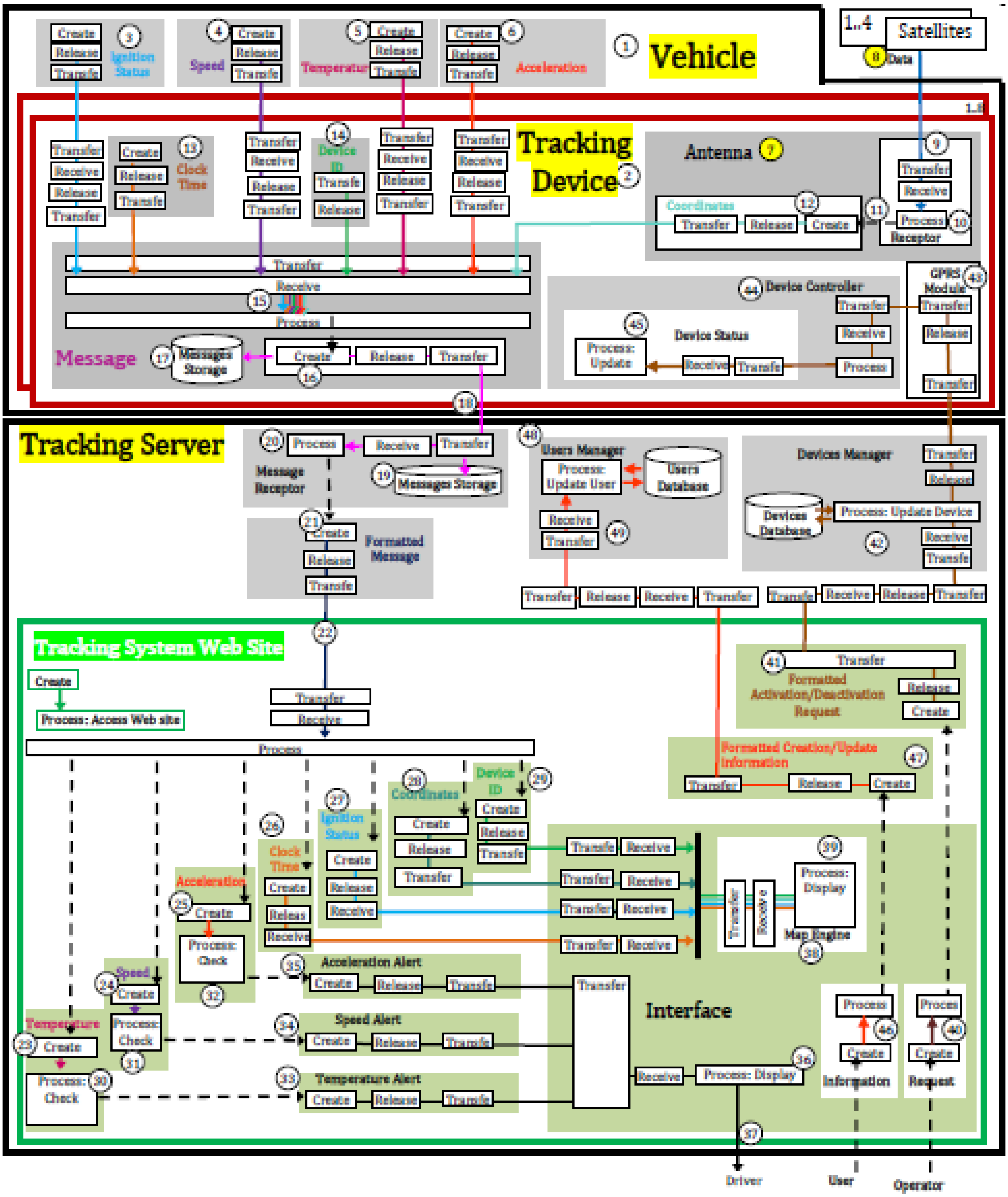

Fig. 12. TM Static Description of the Tracking System.

*b) Plotting updates on a map*: For the purpose of plotting updates on a map, clock time (26), ignition status (27), coordinates (28) and device ID (29) are supplied to the map engine (38) to plot the corresponding update on a map displayed (39) on the interface.

*c) Other activities*: These activities involve sending requests from the user to the server as follows:

- *Device Activation/Deactivation Request*: The operator (40) initiates a request via the interface using a tracking device ID to activate or deactivate the device. The request is formatted by the tracking system (41) and sent to the device manager (42) in the server where the record of the device is updated with the new status. Additionally, the request is sent via a GPRS (43) connection to the corresponding device controller (44) to update the status of the device (45).

- *User Creates/Updates Information*: The user/controller can supply the interface with information and privileges to be created and updated (46). The information is formatted (47) and flows to the user manager in the tracking server (48) to update the user database (49) either by creating a new record or updating an existing record.

## C. *Description of the Dynamic System*

Now, based on the static description, we develop the set of meaningful events in the tracking system in Fig. 12. These events are as follows (see Fig. 14):

Event 1 ($E_1$): Parameters are originated from the vehicle.
Event 2 ($E_2$): The tracking device creates data.
Event 3 ($E_3$): Satellites send data.
Event 4 ($E_4$): Satellite data are received.
Event 5 ($E_5$): Coordinates are calculated.
Event 6 ($E_6$): Tracking device message is created.
Event 7 ($E_7$): Tracking message arrives to the server.
Event 8 ($E_8$): The data of the message are formatted.
Event 9 ($E_9$): The tracking system is operational.
Event 10 ($E_{10}$): The message arrives to the tracking system.
Event 11 ($E_{11}$): Time, ignition status, coordinates and device ID are used by the map engine.
Event 12 ($E_{12}$): Temperature, speed and acceleration generate alerts that are displayed and sent to the driver.
Event 13 ($E_{13}$): The user feeds new user information.
Event 14 ($E_{14}$): The operator requests updating the device status.
Event 15 ($E_{15}$): The server updates new information.
Event 16 ($E_{16}$): The server receives the operator's request to update device status.
Event 17 ($E_{17}$): The device receives the new status.
Event 18 ($E_{18}$): The device updates the status.

Fig. 15 shows the chronology of events in Fig. 14.

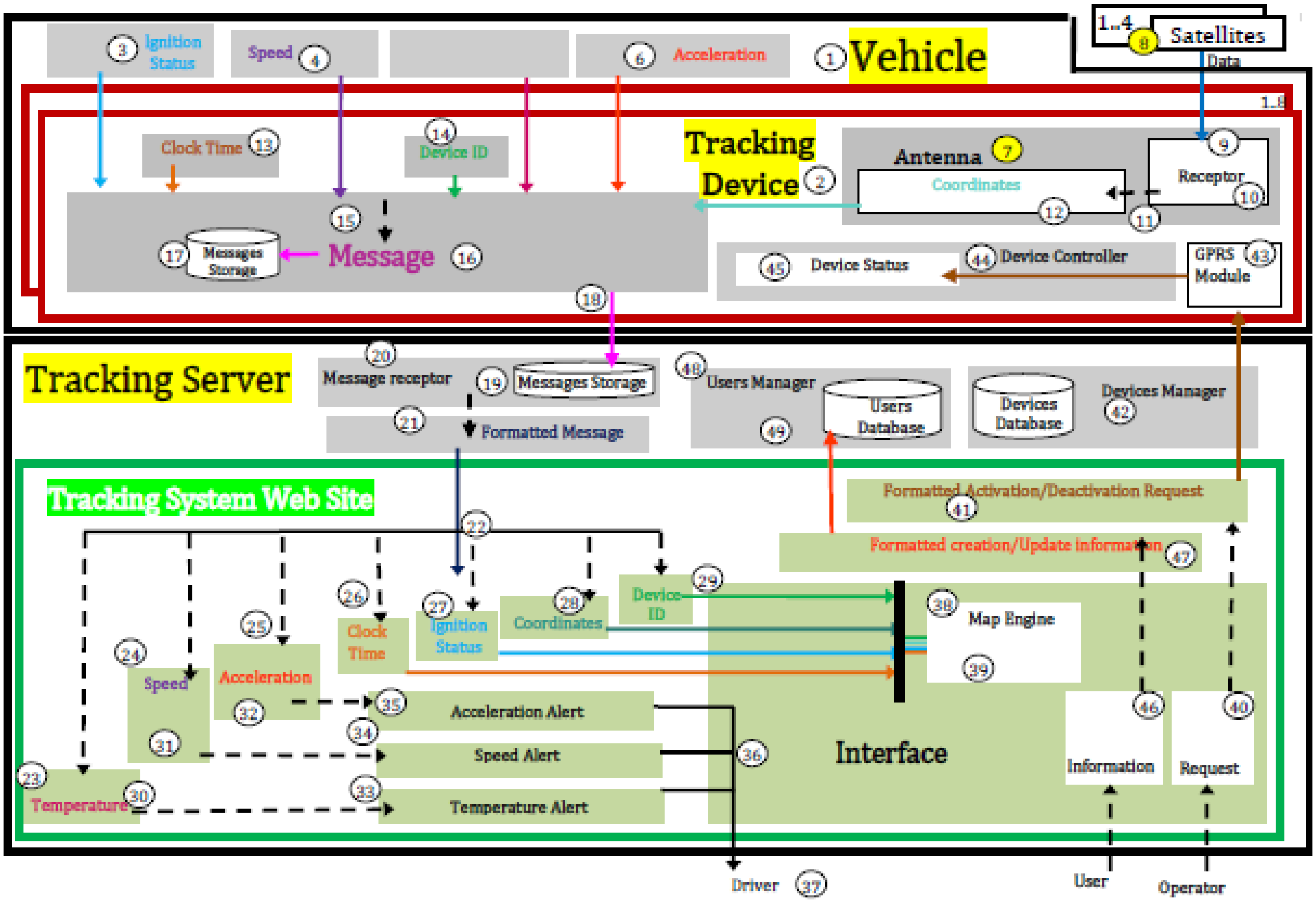

Fig. 13. Simplification of the TM Static Description of the Tracking System Shown in Fig. 12.

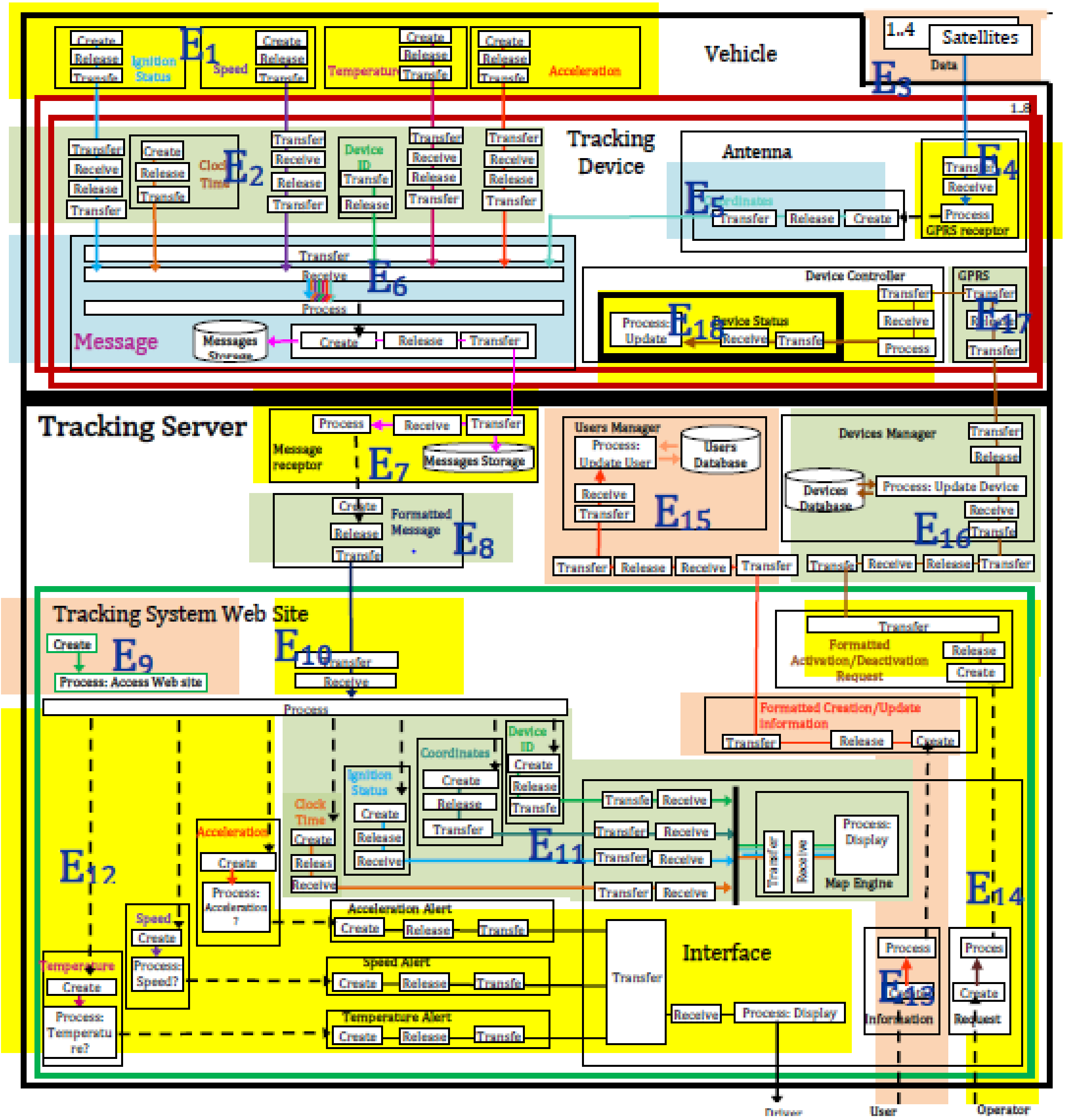

Fig. 14. The Events of the Tracking System.

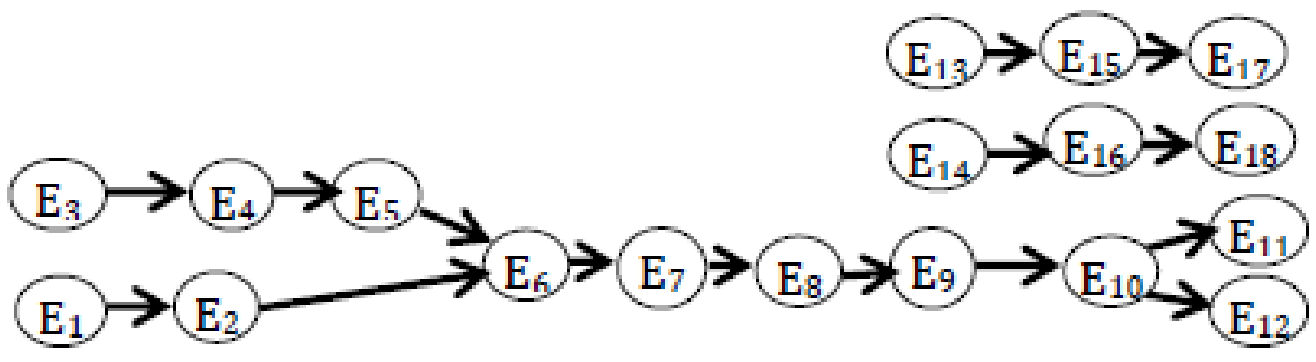

Fig. 15. The Chronology of Events in Fig. 14.

## VI. Conclusion

We proposed using a new modeling technique, TM, to describe a tracking system systematically. Using the TM modeling technique as a documentation tool is a promising field that needs further in-depth exploration to develop appropriate diagramming tools.

The TM diagrams may look complex; however, as previously described, they can be simplified by lumping the details together or consolidating stages. Nevertheless, the underlying TM schema remains a reference for any simplified representation or other uses such as analysis and documentation. Many issues remain to be clarified; however, this paper demonstrates the potential feasibility of this documentation approach, which can be utilized in technical manuals.

The tracking system described in the previous TM diagrams is a viable general solution for tracking by itself; however, it can be the foundation for building larger and more sophisticated advanced fleet management and optimization systems. An example of such utilization would be a customized system dedicated to managing public transportation. Such a system uses the TM modeling methodology to integrate equipment and to facilitate the management and logistical operations for public transportation.

An RFID passenger counter can be integrated with the tracking server through the tracking device to automate the billing and accounting, or fatigue-detecting cameras managed by the tracking device can be used.

Another utilization of the TM model is managing a fleet of self-driving vehicles. In this model, the driver element is eliminated, but instead the tracking device monitors the vehicle. Security checks are done to make sure a vehicle is acting according to its intended driving pattern. The TM model can be applied to ethical guidelines with respect to the three laws of robotics:

- A robot may not injure a human being or, through inaction, allow a human being to come to harm.
- A robot must obey the orders given to it by human beings except where such orders would conflict with the First Law.
- A robot must protect its own existence as long as such protection does not conflict with the First or Second Laws.

Additional controls that can be modeled include continuously assigning the tracking device to check that all the vehicle functionalities are at an acceptable condition. If a problem is detected, the device triggers the movement of the vehicle to the nearest workshop that can solve the detected problem.